\documentclass{aa}
\usepackage{natbib}
\bibpunct{(}{)}{;}{a}{}{,}
\usepackage{txfonts}
\usepackage{graphicx}

\title{Identification of a new short-period comet near the sun}

\author{Sebastian F. H\"onig}
\institute{Max-Planck-Institut f\"ur Radioastronomie, Auf dem H\"ugel 69, 53121 Bonn, Germany}

\offprints{S. F. H\"onig, \\ \email{shoenig@mpifr-bonn.mpg.de}}
\date{Received 6 April 2005 / Accepted September 2005}

\abstract{We present the identification of comet C/1999 R1 (SOHO) with comet C/2003 R5 (SOHO). Both apparitions were only observed with the Solar and Heliospheric Observatory (SOHO) at distances smaller than 0.1\,AU from the sun with the LASCO coronagraphs onboard the spacecraft. Although SOHO comets usually have poor orbital coverage, the 1999 and 2003 arcs are sufficient to generate a link that seems to satisfy all observations. We also analyze comet C/2002 R5 (SOHO) which has similar orbital elements. A fragmentation scenario is proposed and discussed which would support the linkage of C/1999 R1 and C/2003 R5 and thus its short periodic nature.
		}

\authorrunning{S. F. H\"onig}

\begin{document}

\maketitle

\section{Introduction}

Since 1996 the {\it Solar and Heliospheric Observatory (SOHO)} has regularly observed comets close to the sun with its Large Angle Spectrometric coronagraph (LASCO) C2 and C3 cameras. C2 has a circular field of view with a diameter of around $1.5\degr$ while C3 reaches out to almost $4\degr$ solar elongation. Most of the observed comets belong to the well-known Kreutz comet group \citep{Bie02}. There appeared, however, a number of non-Kreutz comets with perihelion distances $0.01\,\mathrm{AU} < q < 0.1\,\mathrm{AU}$. Similarities in orbital elements of a number of objects led to the establishment of three other groups of near-sun comets \citep{Gre02,Mey03}.

While the Meyer group has a mean inclination $i = 72.44\degr\,$, comets of the Marsden and Kracht groups have rather small inclinations (see Table \ref{tab1}). Shortly after the discovery of these new groups \citet{Mar02} remarked on similarities in the orbital elements of the Marsden group comets and comet 96P/Machholz, which has already been associated with various meteor streams. Discussions on the long-term evolution of the Quadrantids \citep{Wil79} and 96P \citep{Ric88} have been brought together by \citet{McI90}, who included the Daytime Arietids as well as the Southern and perhaps Northern $\delta$ Aquarids into this complex of meteor streams and a comet. More recently it was suspected that near-earth asteroid 2003 EH1 is the parent body of the Quadrantids \citep{Jen03} and that C/1490 Y1, the Ursids, Carinids and $\kappa\,$ Velids might also fit into the scheme \citep{McI90,Bab92,Oht03,Sek05}. Given that the Kracht and Marsden groups fit into this complex with short periodic components in the range of 4 to 6 years and their low inclination, attempts have been made to find identification among the comet groups. In fact, two pairs of Marsden group comets could be found where formal links have been presented by \citet{Mar04b,Mar05}. On the other hand, two suspect Kracht group pairs could not be unequivocally linked due to poor measurments and ambiguities.
\begin{table}
\caption{Mean orbital elements for new comet groups.}\label{tab1}
\centering
\vspace{0.1cm}
\begin{tabular}{c c c c}
\hline\hline
& Meyer group & Marsden group & Kracht group \\ \hline
$q$ [AU]& 0.0359 & 0.0481 & 0.0449 \\
$\omega$ [deg] & 57.26 & 22.65 & 59.38 \\
$\Omega$ [deg] & 72.98 & 81.92 & 42.78 \\
$i$ [deg] & 72.44 & 27.14 & 13.25 \\ \hline
\multicolumn{4}{l}{$q$...Perihelion distance, $\omega$...Argument of perihelion,}\\
\multicolumn{4}{l}{$\Omega$...Ascending node, $i$...Inclination}
\end{tabular}
\end{table}

Further to the comets of the Marsden and the Kracht groups there exist another three SOHO comets with similar low-$i$ orbital elements which do not fit into the Machholz interplanetary complex. These are C/1999 R1 (SOHO), C/2002 R5 (SOHO) and C/2003 R5 (SOHO). Remarks on the similarity of the apparent C2 tracks of C/1999 R1 and C/2002 R5 came from R. Kracht shortly after discovery. He also noted the possible relation of C/2003 R5 to this pair.
\begin{table*}
\caption{Parabolic orbital elements for the individual apparitions from C2 observations only which has been used as a starting point for the link, in comparison to the actual C2 link. For the elements two data points from 2003 with overall residuals larger than 50 arcseconds have been excluded.}\label{link}
\centering
\vspace{0.1cm}
\begin{tabular}{l c c c c c}
\hline\hline
& C/1999 R1 (SOHO) &  & C/2003 R5 (SOHO) & \hspace{0.3cm} & C/1999 R1 = C/2003 R5 (SOHO) \\ \cline{2-2} \cline{4-4} \cline{6-6}
Epoch & 1999 Sep 5.0 & & 2003 Sep 8.0 & & 2003 Sep 8.0 \\
$T$ & 1999 Sep 5.53 & & 2003 Sep 8.82 & & 2003 Sep 8.82 \\
$q$ [AU] & 0.0556 & & 0.0569 & & 0.0570 \\
$e$ & 1.0 & & 1.0 & & 0.97757\\
$\omega$ [deg] & 43.88 & & 44.04 & & 43.58\\
$\Omega$ [deg] & 5.65 & & 4.81 & & 5.04 \\
$i$ [deg] & 13.60 & & 14.14 & & 13.67 \\
$P$ [yr] & n/a & &n/a & & 4.01 \\
RMS [\arcsec] & 5.4 & & 13.7 & & 13.6 \\ \hline
\multicolumn{6}{l}{$T$...Time of perihelion, $q$...Perihelion distance, $e$...Eccentricity, $\omega$...Argument of perihelion,} \\
\multicolumn{6}{l}{$\Omega$...Ascending node, $i$...Inclination, $P$...Orbital period, RMS...Root mean square of residuals}
\end{tabular}
\end{table*}
\section{Observations and initial orbits}

Astrometric positions and initial parabolic orbital elements for C/1999 R1 (SOHO), C/2002 R5 (SOHO) and C/2003 R5 (SOHO) have been published in {\it MPEC 1999--R19}, {\it MPEC 2002--S35} and {\it MPEC 2004--J59}, respectively. First indications about similarities of C/1999 R1 and C/2003 R5 come from their appearance in SOHO images: Both cover the same orbital arc, have same apparent tracks and very similar brightness evolution during their passages though C2 and C3, a fact noted by R. Kracht shortly after discovery. They showed no sign of degrading or break-up. On the other hand, C/2002 R5 was less bright which results in a shorter orbital arc. Its apparent track, however, was quite similar to the other comets and led R. Kracht to the conclusion that they may form the core of another new comet group \citep{Mar04a}.

\section{Identification C/1999 R1 (SOHO) = C/2003 R5 (SOHO)}

Orbital calaculations were accomplished using Bill Gray's {\it Find\_Orb} software\footnote{see {\it http:$//$www.projectpluto.com/find\_orb.htm}}. It includes perturbations by all major planets plus the Earth's moon and the three largest minor planets (1) Ceres, (2) Pallas and (4) Vesta. Additionally, relativistic effects due to the comet's small perihelion distance are considered in the software.

To get a meaningful link a first step is to choose the most validated observations. The main problem is that C3 has a scale of 55\arcsec/pixel, while C2 has 12\arcsec/pixel. For the initial orbital calculation we thus chose only C2 observations. It has to be noted that the 2003 data show higher individual residuals than the 1999 observations, regardless of whether they are taken as the starting point or not. For the calculations initial parabolic elements for one apparition were taken and the eccentricity $e$ was forced also to match the other apparition. After a rough fit this constraint was removed and further integration steps led to convergence of the final link. Table \ref{link} presents elements for C/1999 R1 (SOHO) and C/2003 R5 (SOHO) from C2 observations only. They have been used as a starting point for the linking, as well as the actual C2 link. Additionally, a link with both C2 and C3 data was calculated (Table \ref{tab2}) where individual C2 observations are weighted 6 times more than individual C3 observations. Considering that there are around 3 times as many C3 positions as for C2 the latter ones are overweighted by roughly a factor of two -- although C3 still covers a much longer orbital arc. A residual table is available online in the A\&A version. Table \ref{tab2} shows predictions for the next perihelion passage calculated from the C2 double weighted elements. We want to note that the whole solution appears to be stable in the sense that use of shorter integration steps does not lead to other elements with smaller residuals or residuals with similar quality. These can be taken as an indication that the link may be correct. Actually a comet with a period of around 4 years is somewhat unusual since there is only comet 2P/Encke with a shorter period. Further arguments for the periodicity of the object, however, will be given in Sect. \ref{roleof02R5} and the nature of the object -- comet or asteroid -- discussed in Sect. \ref{comast}.
\begin{table}
\caption{Orbital elements of the C2 double weighted link (see text) for the 2003 apparition, as well as a prediction for the next perihelion passage in 2007.}\label{tab2}
\centering
\vspace{0.1cm}
\begin{tabular}{l c c c}
\hline\hline
& \multicolumn{3}{c}{C/1999 R1 = C/2003 R5 (SOHO)} \\ \cline{2-2} \cline{4-4}
Epoch & 2003 Sep 9.0 & & 2007 Sep 11.0 \\
$T$ & 2003 Sep 8.819 & & 2007 Sep 11.263 \\
$q$ [AU] & 0.0570 & & 0.0538 \\
$e$ & 0.97741 & & 0.97860 \\
$\omega$ [deg] & 43.713 & & 48.612 \\
$\Omega$ [deg] & 5.022 & & 0.019 \\
$i$ [deg] & 13.687 & & 12.727 \\
$P$ [yr] & 4.01 & & 3.99 \\
RMS [\arcsec] & \multicolumn{3}{c}{28.213}\\ \hline
\end{tabular}
\end{table}
\begin{figure*}
\centering
\includegraphics[angle=0,width=17cm]{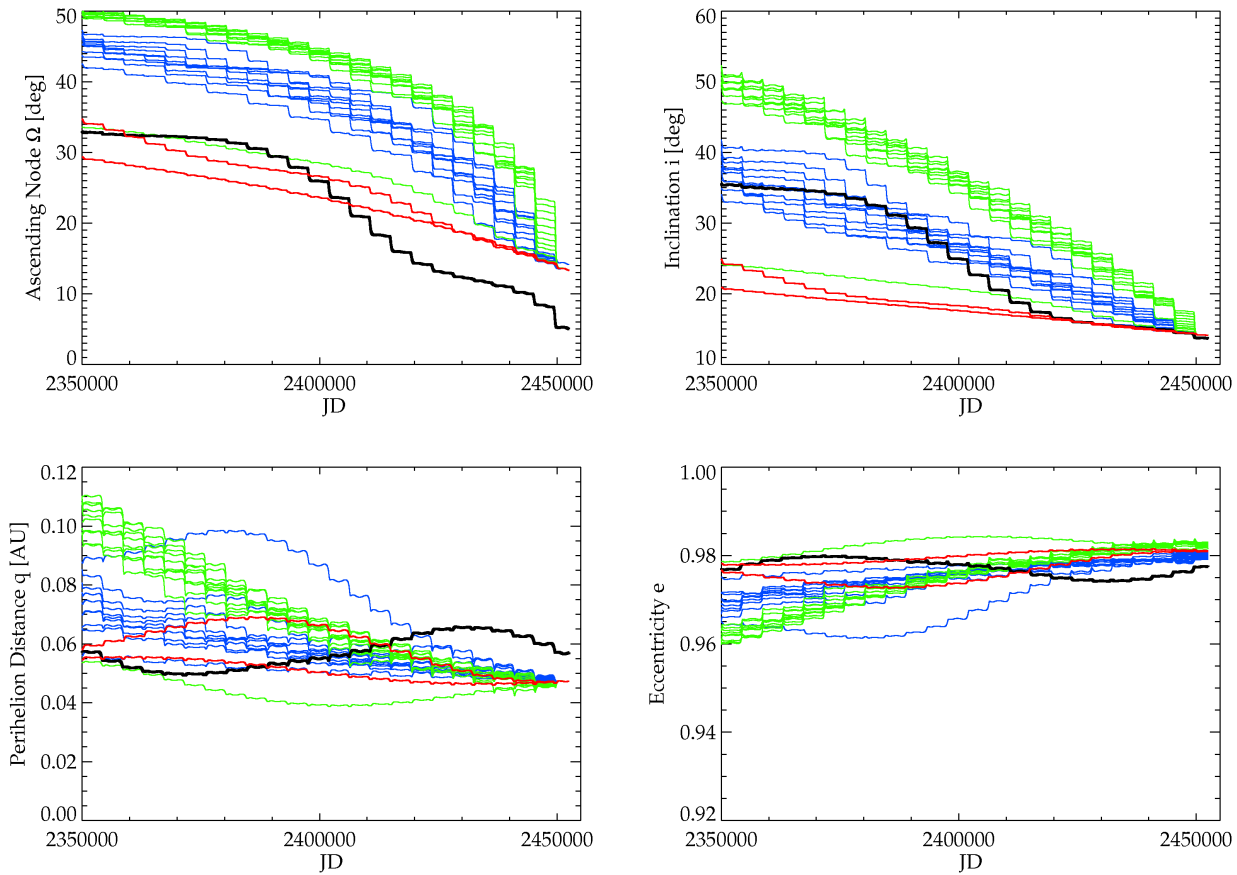}
\caption{Backwards integration of various possible orbital elements for C/2002 R5 (SOHO). Each line represents one set of elements with a certain $3.40\,\mathrm{yr} < P < 4.65\,\mathrm{yr}$, according to the results for C/1999 R1 = 2003 R5 (SOHO). The black line indicates the orbital evolution of C/1999 R1 = C/2003 R5 (SOHO), green and blue lines represent elements with $P < 3.90\,\mathrm{yr}$ and $P > 3.95\,\mathrm{yr}$, respectively. The red lines mark the elements for $P = 3.90\,\mathrm{yr}$ and $P = 3.95\,\mathrm{yr}$ which show best overall agreement with C/1999 R1 = C/2003 R5 (SOHO) (see text). The x-axis of each panel represents Julian Date (JD), while the y-axis represents the Ascending Node $\Omega$, Inclination $i$, Perihelion Distance $q$ and Eccentricity $e$, respectively.}\label{fig1}
\end{figure*}

\section{The role of C/2002 R5 (SOHO)}\label{roleof02R5}

Since C/2002 R5 (SOHO) has similar orbital elements to C/1999 R1 (SOHO) = C/2003 R5 (SOHO) it is worth considering whether there may be some physical association. Calculations show that a direct linkage with a periodicity of $P \sim 3.0\,$yr to C/1999 R1 is not possible due to large overall residuals and systematic trends therein. The same applies to the 2003 apparition and C/2002 R5. Together with its difference in brightness compared to C/1999 R1 and C/2003 R5 we can rule out that it is identical with one of the latter two objects. In view of the fragmentation history of some members of the other low-inclination comet groups \citep{Mar05} C/1999 R1 = C/2003 R5 (SOHO) and C/2002 R5 (SOHO) may have a common origin. In other words: C/2002 R5 (SOHO) may have separated from the parent body C/1999 R1 = C/2003 R5 (SOHO) sometime in the past -- which seems plausible due to the potential parent body's recurring close approaches to the sun. Therefore, it is highly interesting to perform backwards integration of both objects to see if the orbital elements have been more identical at a certain time in the past. For this purpose we have to assume that C/2002 R5 is also of short periodic nature. If it is really possible to find hints for a common origin within this scenario, it would be a strong argument for the periodicity of C/1999 R1 = C/2003 R5 (SOHO) aside from the linkage.

As a starting point we assume that C/2002 R5 has somewhat similar periodicity to C/1999 R1 = C/2003 R5. This can be satisfied by the assumption of fragmentation. We did, however, take a broad range of orbital elements with $3.40\,\mathrm{yr} < P < 4.65\,\mathrm{yr}$ in steps of 0.05 yr for the backwards integration. A set of orbital elements for each step was calculated and put into the freely available orbital integrator {\it Solex 8.5}\footnote{see {\it http:$//$chemistry.unina.it/$\sim$alvitagl/solex/}}, developed by \citet{Vit97}. All 26 solutions for C/2002 R5, together with the orbital elements for C/1999 R1 = C/2003 R5 have
\begin{table}
\caption{Comparison of the orbital elements of C/2002 R5 (SOHO) with a initial periodicity of $P = 3.90\,\mathrm{yr}$ to C/1999 R1 = C/2003 R5 (SOHO) at around the assumed separation time in the late 19th century.}\label{tab21}
\centering
\vspace{0.1cm}
\begin{tabular}{l c c c}
\hline\hline
& C/1999 R1 = C/2003 R5 & & C/2002 R5 (SOHO) \\ \cline{2-2} \cline{4-4}
$q$ [AU] & 0.060 & & 0.060 $\pm$ 0.006 \\
$e$ & 0.976 & & 0.976 $\pm$ 0.002\\
$a$ [AU] & 2.52 & & 2.51 $\pm$ 0.25\\
$\omega$ [deg] & 30.3 & & 34.6 $\pm$ 3.0 \\
$\Omega$ [deg] & 18.3 & & 24.2 $\pm$ 3.2 \\
$i$ [deg] & 18.8 & & 17.4 $\pm$ 1.1 \\ \hline
\end{tabular}
\end{table}
been integrated backwards simultaneously. In general, orbits for SOHO comets show large RMS residuals due to non-perfect astrometrical circumstances. Thus backwards integration is a problematic issue, especially in the case of C/2002 R5 where there is only a small orbital arc available and the data quality is not the best. Since we have the $P$-constraint from C/1999 R1's orbit we think that integrating the above mentioned set of elements will give us a rough idea of the object's past 200\,000 days ($\sim$ 544 yr), although no final statement on the object's past will be possible.

Our calculations illustrate that the orbital developments of C/1999 R1 = C/2003 R5 and C/2002 R5 are quite different within the simulated time span (see Fig. \ref{fig1}). Especially the inclination evolution is quite extreme for most of the elements of C/2002 R5 in comparision to C/1999 R1 = C/2003 R5. There is, however, a certain time span around 2410000 JD where the elements have been quite similar for both objects for periodic solutions around the interval $3.90\,\mathrm{yr}\,<P<3.95\,\mathrm{yr}$. There is no other time within the last 550 yr where such a good agreement in orbital elements can be achieved. Around 2410000 JD a good match of the Argument of Perihelion  $\omega$, the inclination $i$, the eccentricity $e$, the perihelion distance $q$, and thus also the semi-major axis $a$ to C/1999 R1 = C/2003 R5 is found. Furthermore the difference in Longitude of Ascending Node $\Omega$ is smaller than 6\degr\,(see Table \ref{tab21}). Although the orbits of C/2002 R5 and C/1999 R1 = C/2003 R5 seem to be close together in the late 18th century, the difference in the angular elements did hardly improve with respect to the present elements. Additionally, the longitude of perihelion of C/2002 R5 ($L=58\degr$) is still way off the value for C/1999 R1 = C/2003 R5 ($L=48\degr$), while the latitude of perihelion matches quite well ($B=9\degr$ for both objects). This may be an effect of the unknown scatter in initial conditions for C/2002 R5. As a measure for this scatter we illustrated the uncertainities of the elements, based on the variations for one periodicity interval $dP=0.05yr$, in Table \ref{tab21}. The resulting deviation may partly explain the large difference in $L$ while the agreement is still not striking. For a final statement on this scenario it is necessary to improve the quality of the current orbit. If C/2002 R5 is really periodic and has seperated from the parent body in the late 19$^{th}$ century (probably around 1890 $\pm$ 10), we may soon reobserve it.

The first step to test this scenario would be a search for C/2002 R5 in LASCO data at the previous return. The problem is that it should have appeared in September or early October, 1998. At around this time there are no LASCO images availbale since SOHO experienced major problems. Therefore we encourage observations at the possible next return in 2006 (see Table \ref{tab3}). On the other hand, periodicities which placed the object around perihelion later than the end of October 1998 or earlier than July 1998 can almost certainly be ruled out since there was nothing found in the LASCO archives, although thoroughly inspected by several people. It has to be pointed out that the actual time of perihelion for the next return is much more uncertain than the difference in the orbits may suggest.

\begin{table}
\caption{Predicted orbital elements for a possible next return of C/2002 R5 (SOHO), based on $P = 3.90\,\mathrm{yr}$ and $P = 3.95\,\mathrm{yr}$ for the 2002 apparition, respectively.}\label{tab3}
\centering
\vspace{0.1cm}
\begin{tabular}{l c c}
\hline\hline
& \multicolumn{2}{c}{C/2002 R5 (SOHO)} \\ \cline{2-3}
Epoch & 2006 Jul 30.0 & 2006 Aug 17 \\
$T$ & 2006 Jul 30.4 & 2006 Aug 17.7 \\
$q$ [AU] & 0.0470 & 0.0470 \\
$e$ & 0.981 & 0.981 \\
$\omega$ [deg] & 45.92 & 45.94 \\
$\Omega$ [deg] & 13.15 & 13.13 \\
$i$ [deg] & 14.10 & 14.09 \\
$P$ [yr] & 3.90 & 3.95\\ \hline
\end{tabular}
\end{table}

\section{Comet or minor planet?}\label{comast}

\begin{table*}
\caption{Determination of absolute magnitude from photometric data of C/1999 R1 (SOHO).}\label{tab4}
\centering
\vspace{0.1cm}
\begin{tabular}{c c c c c c c c}
\hline\hline
& Magnitude & \multicolumn{2}{c}{Distance} & Phase Angle& & \multicolumn{2}{c}{Absolute Magnitude} \\
& $m$ [mag] & $r$ (helioc.) [AU] & $d$ (geoc.) [AU]& $\phi$ [deg] & & cometary $H_C$ [mag] & asteroidal $H_{MP}$ [mag]\\ \hline
& 6.6 & 0.0577& 0.9978& 116.9 & & 12.84 & 12.84 \\
& 6.5 & 0.0573& 0.9833& 110.9 & & 12.75 & 12.75 \\
& 6.4 & 0.0572& 0.9870& 106.6 & & 12.64 & 12.64 \\
& 6.3 & 0.0573& 0.9907& 102.3 & & 12.53 & 12.53 \\
& 6.2 & 0.0575& 0.9945& 98.1 & & 12.41 & 12.41 \\
& 6.2 & 0.0580& 0.9993& 93.3 & & 12.38 & 12.38 \\
& 6.4 & 0.0592& 1.0071& 85.9 & & 12.52 & 12.52 \\
& 6.4 & 0.0600& 1.0109& 82.1 & & 12.49 & 12.49 \\
& 6.7 & 0.0609& 1.0148& 78.4 & & 12.75 & 12.75 \\
& 6.9 & 0.0619& 1.0186& 74.9 & & 12.90 & 12.90 \\
& 6.9 & 0.0645& 1.0272& 68.2 & & 12.79 & 12.80 \\
& 7.4 & 0.0676& 1.0355& 62.2 & & 13.17 & 13.20 \\
& 7.4 & 0.0706& 1.0427& 67.7 & & 13.07 & 13.11 \\ \hline
\end{tabular}
\end{table*}

It is widely known that almost no SOHO comet displays a tail, except for a number of Kreutz group members and very few bright other comets. Nevertheless, all objects are designated and referred to as comets, mostly because of the parabolic elements that fit into this scheme, and the origin of the Kreutz group where small tail-less members are fragments of well-observed cometary progenitors. The suggested connection between comet 96P/Machholz and the Kracht and Marsden groups supports this procedure. Additionally there are hints that the identified recurring objects of these groups show a trend to have fainter second perihelion passages as a result of their losing a lot of material and thus desintegrate over time \citep{Sek05}.

Given the semi-major axis $a \sim 2.52\,\mathrm{AU}\,$ of C/1999 R1 = C/2003 R5, one may ask whether this object is really a comet or an Alinda-type asteroid close to the 3:1 resonance with Jupiter. In fact, no tail was visible, in either C3 or in C2 images. It is also not clear if the object displayed a coma. We accomplished an analysis of photometric data acquired during the 1999 perihelion passage \citep{Bie99} and searched for possible phase effects which would be typical asteroidal behaviour. It has to be noted that photometric measurements from LASCO images are highly affected by a number of disturbing factors, such as severe radial vignetting or obscuration by the pylon \citep{Bie02}. An analysis of the absolute magnitude of this object with standard minor planet ($G=0.15$) and cometary ($n=2.0$) slope parameters give no conclusive answer (Table \ref{tab4}). Both possibilities satisfy the photometric data and allow for similar standard deviation in their absolute magnitudes. It has to be remarked that the absolute value for $H_{MP}$ is not consistent with the standard translation to the object's size ($\sim10\,\rm{km}$). From estimations for other SOHO comets its size should be of the order of 10s of meters up to perhaps 100 meters, depending on its actual nature.

\section{Discussion}

We present the identification of a new short periodic comet which was only observed with the SOHO spacecraft so far. We linked the objects C/1999 R1 and C/2003 R5 as two consecutive perihelion passages of the same object. Orbital elements for the next return in 2007 were calculated. Additionally, we show that the third object, C/2002 R5, with similar orbital elements may be a fragment of the periodic comet and has separated from it some time in the late 19$^{th}$ century. Aside from the orbital link this supports the idea of short periodicity of C/1999 R1 = C/2003 R5. If the whole scenario is true, C/2002 R5 may be observed in 2006 with the SOHO satellite. In the last section we discussed the nature of C/1999 R1 = C/2003 R5 and concluded that from currently available photometric measurements it is impossible to tell if it is a comet or a minor planet. With this identification the object should also become accessible to ground-based observations, so that the situation may improve soon. It is, therefore, strongly encouraged to try ground-based recovery of the object within the observable window just before the perihelion approach in 2007.

\bibliographystyle{aa}

\label{lastpage}

\end{document}